\documentclass[pra,twocolumn,amsmath,amssymb,byrevtex]{revtex4}

\usepackage{bm,amssymb,amsmath,graphicx,color}

\newtheorem{Thm}{Theorem}
\newtheorem{Prop}{Proposition}
\newtheorem{Lem}{Lemma}

\newtheorem{Def}{Definition}

\newcommand{\tr}{\mathop{\mathrm{tr}}\nolimits}

\newcommand{\HA}{\mathop{\mathcal{H}}\nolimits}

\newcommand{\LA}{\mathop{\mathcal{L}}\nolimits}
\newcommand{\B}{\mathop{\mathcal{B}}\nolimits}

\newcommand{\SA}{\mathop{\mathcal{S}}\nolimits}
\newcommand{\R}{\mathop{\mathbb{R}}\nolimits}\newcommand{\N}{\mathop{\mathbb{N}}\nolimits}\newcommand{\I}{\mathop{\mathbb{I}}\nolimits}\newcommand{\A}{\mathop{\mathcal{A}}\nolimits}

\newcommand{\E}{\mathop{\mathcal{E}}\nolimits}

\newcommand{\F}{\mathop{\mathcal{F}}\nolimits}

\newcommand{\ketbra}[2]{| #1 \rangle \langle #2 |}
\begin{document}

\title{Physical Equivalence of Pure States and Derivation of Qubit in General Probabilistic Theories}
\author{Gen Kimura ${}^{[a,b]}$}
\email{gen-kimura[at]aist.go.jp}
\author{Koji Nuida ${}^{[b]}$}
\email{k.nuida[at]aist.go.jp}
\author{Hideki Imai ${}^{[b],[c]}$}
\affiliation{[a] Research and Development Initiative, Chuo University, 1-
13-27 Kasuga, Bunkyo-ku, Tokyo 112-8551, Japan}
\affiliation{[b] Research Center for Information Security (RCIS),
National Institute of Advanced Industrial
Science and Technology (AIST). 
Daibiru building 1003,
1-18-13 Sotokanda, Chiyoda-ku, Tokyo, 101-0021, Japan \\ 
[c] Graduate School of Science and Engineering,
Chuo University.
1-13-27 Kasuga, Bunkyo-ku, Tokyo 112-8551, Japan
}


\begin{abstract}
In this paper, we investigate a characterization of Quantum Mechanics by two physical principles based on general probabilistic theories. 
We first give the operationally motivated definition of the physical equivalence of states and consider the principle of the physical equivalence of pure states, 
which turns out to be equivalent to the symmetric structure of the state space. 
We further consider another principle of the decomposability with distinguishable pure states. 
We give classification theorems of the state spaces for each principle, and derive the Bloch ball in $2$ and $3$ dimensional systems by these principles. 
\end{abstract}
\maketitle

\section{Introduction}

After the completion of von Neumann's celebrated axioms \cite{ref:vN} of Quantum Mechanics (QM), one of the theoretical and philosophical interests of QM shifted to derive the theory using solely physical principles. 
Here the physical principle means a physical statement which is, in principle, testable by experiments. 
For instance, a statement such as ``a quantum observable is represented by a self-adjoint operator on a Hilbert space'' is not a physical principle, 
but a statement such as ``the speed of light is constant independent of observer's motion'' gives a typical example of a physical principle. 
Indeed, the standard axioms of QM presuppose a priori mathematical structures such as Hilbert space and linear operators --- as the former example is one of the axioms --- 
and then gives a recipe (Born's rule) of how to predict physical phenomena by combining such mathematical objects. 
This non-physical characterization of QM is sometimes considered as one of the causes why QM is difficult to intuitively understand
 --- in the end, many students who learns QM for the first time naively ask themselves ``Why observables are described by (non-commutative) operators?''
On the other hand, if QM was constructed solely with physical principles and if, moreover, they are tested by experiments, then 
QM would be intuitively more acceptable than the present situation, as all the ``weird" quantum phenomena (such as the uncertainty principle and an entanglement) automatically follow from experimentally established phenomena. 
(For instance, most physicists feel to intuitively understand the relativity of time, even though contrary to our common sense, 
since they know that the invariance of the speed of light is experimentally confirmed and the fact naturally leads the relativity of time.)

Therefore, a desire to derive axioms of QM which are purely described by physical principles is natural as many researchers indeed have tackled this problem so far \cite{ref:qlog,ref:algebra,ref:Ludwig,ref:Fuchs02,ref:Clifton,ref:Dariano}, and some of recent developments in \cite{ref:Hardy,ref:Masanes,ref:DakicBrukner} and \cite{ref:CDP}  contribute the goal in each ways.

One of the lines of this research starts by recognizing that QM is one of the probabilistic theories. 
Indeed, with the standard interpretation of QM and indeed with a practical method to compare with experimental data, what QM predicts is a probability to obtain an outcome by performing a measurement under a given state. 
However, the structure of the theory is more complicated than a classical probability theory (based on Kolmogorov's axioms).  
Therefore, it is convenient to first establish a general framework for all the possible probabilistic models including both classical theory and QM. 
Then, we can look for particular conditions (described by physical principles) which narrow down the general models to be QM. 
Indeed, such a general framework has been well investigated (for instance see \cite{ref:Mackey,ref:Araki,ref:Gudder1,ref:Ozawa,ref:HolevoSM}) and recently is called General Probabilistic Theories (GPTs) \cite{ref:Barrett,ref:NKM,ref:KNI}. 

For this purpose, we first give an operationally motivated definition of the physical equivalence between states and propose a physical principle of [P5] {\it Physical Equivalence of Pure States}. 
Then we show that the principle is equivalently described by a symmetric structure of state space (See Theorem \ref{thm:PESym} and the principle [P5']). 
Next, we consider another principle of [P6] the decomposability with distinguishable pure states.  
We give representation theorems for state spaces for each principle (See Proposition \ref{Prop:IF}, Theorems \ref{thm:CSym2}, \ref{thm:CSym3} for [P5], and Theorem \ref{thm:P6} for [P6]). 
By combining these results, we derive that GPTs with $2$ or $3$ dimensional state spaces are either classical or quantum (See Theorem \ref{thm:MAIN}). 
Our results are closely related to those in \cite{ref:Hardy,ref:Masanes,ref:DakicBrukner} and \cite{ref:CDP}, where the general Bloch ball is obtained. 
However, the definition of the Bloch ball there is given by a system with at most two distinguishable states. 
On the other hand, we do not require the maximum number of distinguishable states and discuss how the state space is characterized. 
To understand this non-triviality, we notice that a state space is not uniquely determined under the above two principles but on the contrary still there admits a complicated state space, 
where the typical example is given by QM itself with more than or equal to $3$ dimensional Hilbert spaces (see \cite{ref:Bloch}). 
Notice that in this paper we do not deal with a dynamics, a measurement process, nor a composite system, which are another important notions for GPTs. 
Rather, we are interested in how GPTs are characterized with the above minimum restrictions (principles) posed only on states and measurements.

This paper is organized as follows. 
In Sec. \ref{sec:GPTs}, we review and reconstruct GPTs with natural physical principles [P1-P3], especially explaining the basic notions of states and measurements and how mathematical structures such as vectors in a real vector space appear to represent them. 
In Sec. \ref{sec:AP}, we introduce and investigate the principles [P5] and [P6], including general discussion of the invariant states under any affine bijection on the state space. 
In Sec. \ref{sec:CD}, we give a concluding remark and discussion. 
 
\section{Review of General Probabilistic Theories}\label{sec:GPTs}

In this section, we briefly review General Probabilistic Theories (GPTs) as a general framework of operationally well-defined probabilistic models, focusing on the notions of states and measurements.  
Based on operationally motivated principles, we explain how states and measurements are mathematically represented by vectors in a compact convex set and effects, respectively. 
It is important to recognize that these mathematical structures are not introduced a priori but they are the consequences from physical principles. 

\subsection{States and Measurements on GPTs	}\label{sec:GPT}

The basic ingredients of GPTs are the notions of {\it states} and {\it measurements} with a physical law (e.g., Born's Rule in quantum mechanics) to predict a probability to obtain a measurement outcome. 
Formally, we assume the following:  

 [P1] (Probability Assignment Principle) {\it A probability $\mathrm{Pr}\{M = m_i | s \}$ to get an outcome $m_i$ is given when performing a measurement $M$ under a state $s$:
\begin{eqnarray}
\mathrm{Pr}\{M = m_i | s \} \ge 0, \sum_{i=1}^n \mathrm{Pr}\{M = m_i | s \}  = 1.  
\end{eqnarray}
}
In this paper, we assume for mathematical simplicity that each measurement $M$ has a discrete (and finite) set of measurement outcomes $\{m_i \}_{i=1}^n$ with $n \ge 1$ 
\footnote{In general setting, one may assume that each measurement $M$ has a measurable space $(\Omega,{\cal F})$ where $\Omega$ is a sample space for measurement outcomes and $\F$ is a $\sigma $ algebra for measurement events, e.g., $\Omega = \R, \F = \B(\R)$. 
Then, for each event $\Delta \in \F$ and state $s$, a probability $\mathrm{Pr}\{\Delta \in \B(\R) | s\}$ to get an outcome in $\Delta$ is determined.}. 
In particular, $2$-valued measurements with $n=2$ (Yes-No measurements) play a fundamental role as building blocks of any measurements  
(See for instance Chapter 6 in \cite{ref:Kraus}). 
In the following, the symbols $s,s_1,s_2,\cdots$, $M,M_1,M_2,\cdots$ and $m,m_1,m_2,\cdots$ represent states, measurements and measurement outcomes, respectively. 

With an operational view, we naturally identify states $s_1$ and $s_2$ if there are no physical differences between them. 
In our case, the only method to compare physical properties of states is to observe probabilities given through possible measurements. 
Thus we may assume the following: 

[P2-1] (Separation Principles for States) {\it States $s_1$ and $s_2$ are identified iff $\mathrm{Pr}\{M=m_i | s_1\} = \mathrm{Pr}\{M=m_i | s_2\}$ for any measurement $M$ and measurement outcome $m_i$.  }

Indeed, the preliminary notion of ``state'' in an operational view is defined by a preparation of experimental instruments (settings). 
Then, a state is defined by the equivalence class among all the ``states'' having the same physical properties, which in our case are the probability distributions among possible measurements. 
Therefore, one may interpret [P2-1] as one aspect of a natural definition of states. 

The similar argument follows for measurements. 
Namely, we identify measurements if there are no physical differences among them under any states. 
It is also convenient to drop the information of measurement outcome $m_i$ by identifying them just as an $i$th outcome. 
By doing so, we denote by $\mathrm{Pr}\{ M = i | s \} := \mathrm{Pr}\{ M = m_i | s \}$ the probability to get $i$th outcome when performing a measurement $M$ under a state $s$. 
Thus we identify $n$-valued measurements $M$ and $M^\prime$ irrespective of the set of outcome if their statistical properties are identical for each outcome. 
Therefore, we assume the following: 
  
[P2-2] (Separation Principles for Measurements) {\it Measurements $M_1$ and $M_2$ with the same number of outcomes are identified iff 
$\mathrm{Pr}\{M_1= i | s\} = \mathrm{Pr}\{M_2= i | s\}$ for any $i$ and state $s$. }

It is operationally natural to allow a probabilistic mixture of states as one of the procedures to prepare states.  
Namely, if states $s_1$ and $s_2$ can be prepared, a preparation of state $s_1$ with probability $p$ and state $s_2$ with probability $1-p$ can also be prepared.  
With this preparation, the probability for any measurement should satisfy the mixing property due to the union rule of the probability and the definition of the conditional probability. 
Formally, we thus assume the following:  

[P3-1] (Mixing Principle for States) {\it For any state $s_1$, $s_2$ and for any $p \in [0,1]$, there exists a state $s$ satisfying $
\mathrm{Pr}\{M = i | s \} = p \mathrm{Pr}\{M = i | s_1 \} + 
(1-p)  \mathrm{Pr}\{M = i | s_2 \}$ for any measurement $M$ and $i$. 
}

Of course, we implicitly suppose that the above state $s$ can be prepared as a probabilistic mixture of $s_1$ and $s_2$ with probabilities $p$ and $1-p$. 
From [P2-1], it is uniquely determined with given $s_1,s_2$ and $p$, and is denoted by $s= \langle p;s_1,s_2 \rangle$ \cite{ref:Gudder1}. 
Note that the trivial mixtures $s= \langle p;s,s\rangle$ and $s= \langle 1;s,s'\rangle = \langle 0;s',s\rangle$ follow for any state $s,s'$ and $p \in [0,1]$.  
A state is called a {\it pure} state iff there exist no means to prepare it with nontrivial probabilistic mixture. 
Namely, $s$ is a pure state iff $s = \langle p;s_1,s_2\rangle$ for states $s_1,s_2$ and $p \in (0,1)$ implies $s_1 = s_2$. 
Otherwise, a state is called a {\it mixed} state. 
In the following, we denote the set of all states and pure states by $\SA$ and $\SA_{pure}$, respectively. 
$\SA$ is called the {\it state space}.

Similarly, it is operationally legitimate to allow a probabilistic mixture of measurements:  
Namely, if $n$-valued measurements $M_1$ and $M_2$ can be performed, 
then a measurement to perform $M_1$ with probability $p$ and $M_2$ with probability $1-p$ is also possible: 

[P3-2] (Mixing Principle for Measurements) {\it For any $n$-valued measurement $M_1$, $M_2$, and for any $p \in [0,1]$ there exists an $n$-valued measurement $M$ satisfying 
$\mathrm{Pr}\{M = i | s \} = p \mathrm{Pr}\{M_1 = i | s \} + 
(1-p)  \mathrm{Pr}\{M_2 = i | s \}$ for any $i = 1,\ldots,n$ and state $s$. }

From [P2-2], the measurement $M$ with given $M_1,M_2,p$ in [P3-2] is uniquely determined up to arbitrary choice of measurement outcomes. 

So far, we have only assumed natural principles [P1]-[P3] resorting to no {\it a priori} mathematical structures. 
However, it turns out that states and measurements are represented by vectors in a real vector space as follows: 

[{\bf Representation} 1] (i) {\it A state is represented by a vector in a real vector space $V$ such that a convex combination of states $s_1$ and $s_2$ with weight $p$ represents the state $\langle p;s_1,s_2 \rangle$; 
Thus the state space $\SA$ is a convex subset of $V$ and pure states correspond to extreme points of $\SA$. }  
(ii) {\it An $n$-valued measurement $M$ is represented by an $n$-tuple of effects $(e_i)_{i=1}^n$ on $\SA$ such that $\sum_i e_i = u$, meaning that $e_i(s)$ is the probability to get $i$th outcome when performing the measurement $M$ under a state $s$. }

Remind the following mathematical terminologies: a subset $W $ in a real vector space is called a convex subset iff it is closed under any convex combinations. 
A vector $w \in W$ is called an extreme point iff $w$ has no non-trivial convex decompositions in $W$, i.e., $w = pw_1 + (1-p)w _2$ for some $w_1,w_2 \in W$ and $p \in (0,1)$ implies $w= w_1 = w_2$.  
A real functional $f: W \to \R$ is called an affine functional on $W$ iff $f(p w_1 + (1-p)w_2) = p f(w_1) + (1-p)f(w_2)$ for any $w_1,w_2 \in W$ and $p \in [0,1]$. 
An affine functional $e$ on $W$ is called an {\it effect} on $W$ iff the range is in $[0,1]$. 
The unit effect and the zero effect, denoted by $u$ and $0$, are effects defined by $u(s) = 1, 0(s)= 0$ for all $s \in W$. 
For the readers' convenience, we present the precise proof of Representation 1 in Appendix \ref{app:Rep1} (See Theorem \ref{thm:BasicRep}). 

Notice that a representation of states and measurements is of course not unique. 
GPTs with state representations $\SA \subset V$ and $\SA^\prime \subset V^\prime$ in real vector spaces $V,V^\prime$ are equivalent (hereafter {\it affine isomorphic}) iff there exists an affine bijection $\Lambda: \SA \to \SA^\prime$ such that the correspondent of measurement $M = (e_i)_{i=1}^n$ on $\SA$ is given by $M^\prime = (e_i\circ \Lambda^{-1})_{i=1}^n$ on $\SA^\prime$. 
In particular, if the (affine) dimension of $\SA$ is finite, one may represent states as vectors in an Euclidean space.  

In the following, we use the Representation 1 and let $\SA$ be a state space of a vector representation as above. 
We denote by $\A(\SA)$ and $\E(\SA)$ (or simply $\A$ and $\E$) the sets of all the affine functionals and effects on $\SA$. 
It is easy to show that $\A(\SA)$ is a real vector space with point-wise sum and multiplication and $\E(\SA)$ is a convex subset of $\A(\SA)$. 
An extreme point of $\E(\SA)$ is called a decision effect (or a pure effect) \cite{ref:Ludwig,ref:KNI}.

Notice that, while any measurement $M$ is represented by an $n$-tuples $(e_i)_{i=1}^n$ of effects on $\SA$ satisfying $\sum_i e_i = u$, the opposite might not hold in general. 
However, as it is satisfied in both classical and quantum systems (see Sec.~\ref{sec:ex}), in this paper we assume the following postulate:  

[P4] {\it Given a state space $\SA$, any mathematically well-defined measurement on $\SA$ is feasible. Namely, 
for any $n$-tuples of effects $(e_i)_{i=1}^n$ with $\sum_i e_i = u$, there exists the corresponding measurement $M$ such that $M$ is represented by $(e_i)_i$. } 

With this condition, it turns out that only one has to specify is the state space of GPT, since then feasible measurements are automatically given by effects on $\SA$.

In this paper, we assume [P1-P4] as basic principles for operationally well-defined probabilistic models. 

Next, we explain that a natural topology is introduced into a state space. 
Note that any physical measurements are accompanied by a finite (even small) error. 
Thus, it is operationally natural to say that states $s_1$ and $s_2$ are close iff for arbitrary but finite numbers of measurements the probabilities by measuring them under $s_1$ and $s_2$  are within an arbitrary small error. 
This defines the so-called {\it physical topology} \cite{ref:Araki}. 
With this natural topology, we can go further from the Representation 1: 

[{\bf Representation 2}] 
{\it A state is represented by a vector in a locally convex Hausdorff topological vector space $V$ such that the topology is given by the physical topology and --- without loss of generality --- the state space is a compact convex subset of $V$.  
In particular, if $\mathrm{dim} (\SA) =: d < \infty$, a state is represented by a vector in $\R^d$ such that the physical topology is 
the Euclidean topology and $\SA $ is a compact (i.e., closed and bounded) convex subset in $\R^d$. } 

Note that dim $\SA$ is the dimension of the affine hull Aff$(\SA):= \{\sum_i \lambda_i s_i \ | \ w_i \in \SA, \lambda_i \in \R, \sum_i \lambda_i = 1\}$. 
We refer \cite{ref:ltvs} for a (locally convex Hausdorff) topological vector space $V$. 
Remind that if $V$ is finite dimensional, then the topology is unique. 
Thus, if $V = \R^d$, then the unique topology is given by the Euclidean topology of $\R^d$. 
Any non-empty compact convex subset $W$ of $V$ has the non-empty set of extreme points of $W$ and is the closed convex hull of extreme points (Krein-Milman Theorem). 
In finite dimensional case, any vector $w \in W$ has a convex decomposition with finite numbers of extreme points:  
$w = \sum_{i=1}^m p_i w_i$ with extreme points $\{w_i \}_{i}$ and a probability distribution $(p_i)_{i=1}^m$ (see, for instance, Theorem 5.6 in \cite{ref:Lay}). 
Therefore, in any GPTs, there exist enough pure states so that $\SA$ is a closed convex hull of $\SA_{pure}$. 
Physically speaking, any state $s$ is approximately (in the sense of the physical topology) prepared with a probabilistic mixture of finite numbers of pure states.  
If dim Aff$(\SA) < \infty$, any state $s$ is exactly prepared with a probabilistic mixture of finite numbers of pure states (hereafter, a pure-state decomposition). 
For the readers' convenience, Representation 2 is explained in details in the case of finite dimension in Appendix \ref{app:Rep2}. 
Indeed, one can introduce a natural norm on $V$ and all the topological issues can be described with this norm in the finite dimensional case. 
(See \cite{ref:Araki} and \cite{ref:NKM} for infinite dimensional cases.) 

In the following, we restrict ourselves to finite dimensional GPTs where the state space $\SA$ is embedded into a finite dimensional vector space $V$. 
For the underlying vector space, we assume without loss of generality $V : =$ Aff$(\SA)$.  
In particular, one can assume $\SA$ to be a compact (i.e., closed and bounded) convex subset in a Euclidean space $\R^d$ noting that the Euclidean topology coincides with the physical topology.

Notice that, for any state $s$, one can define an affine functional $\Lambda_s: \E \to [0,1]$ by 
$$
\Lambda_s(e):= e(s) \ (\forall e \in \E), 
$$
which satisfies $\Lambda_s(u) = 1$ and $\Lambda_s(0) = 0$. 
From [P2-1], we have $\Lambda_{s_1} = \Lambda_{s_2}$ iff $s_1 = s_2$. 
Therefore, a state is represented by an affine functional $\Lambda:\E \to [0,1]$ satisfying $\Lambda(u) = 1$ and $\Lambda(0) = 0$. (Indeed, in Appendix \ref{app:Rep1}, a vector representation of states is introduced in this way.) 
It is interesting to notice that the converse is also true: 
\begin{Thm}\label{thm:AF}
Let $\Lambda$ be an affine functional on $\E= \E(\SA)$ such that $\Lambda (e) \in [0,1]$ for any $e \in \E$ and $\Lambda(u) = 1, \Lambda(0) = 0$. 
Then, there exists the unique state $s \in \SA$ such that $\Lambda(e) = e(s)$ for any $e \in \E$. 
\end{Thm}
(The proof of this theorem is given in Appendix \ref{sec:A}.)  
Therefore, one can identify states and affine functionals on $\E$ satisfying the above properties. 
In the following, we omit the symbol $\Lambda_s$ and just write
\begin{equation}\label{eq:se=es}
s(e) = e(s),
\end{equation}
meaning that $s(e):= \Lambda_s(e)$.

To complete the framework of GPTs, one still needs another important notions of transformation (dynamics), measurement process \cite{ref:DaviesOzawa}, and especially composition of systems. 
In this paper, we consider GPTs only from the viewpoints of states and measurements by considering how physical principles posed only on states and measurements can narrow down GPTs. 
This means, of course, that there remains a freedom to further specify GPTs by using another principles on dynamcis, composition of systems, etc. \cite{ref:Hardy,ref:Masanes,ref:DakicBrukner}.

\subsection{Examples of GPTs}\label{sec:ex}

The typical examples of GPTs are Classical Probability Theory and QM. 
Here, we briefly review the finite cases for classical theory and QM: 

[Example 1] (Finite Classical Probability Theory) A state of a finite classical probability theory is described by a probability distribution ${\bm p} = (p_1,\ldots,p_c)$ on a sample space $\Omega_c =\{\omega_1,\omega_2,\ldots,\omega_c\}$, and the state space is given by $\SA_{cl}:= \{{\bm p} = (p_1,\ldots,p_c) \ |\  p_i \ge 0, \sum_i p_i = 1 \} \subset \R^c$. 
There exist $c$ numbers of pure states (vertices): ${\bm p}^{(1)} =(1,0,\ldots,0), \ldots,{\bm p}^{(c)} =(0,\ldots,0,1)$. 
Geometrically $\SA_{cl}$ is the $c-1$ dimensional (standard) simplex. \footnote{
A subset $W$ is called a $c$ dimensional simplex if it is a convex hull of affinely independent set $\{w_i\}_{i=1}^{c+1}$. 
Remind that $\{v_i \in V \}_{i=1}^{m+1}$ is called an affinely independent set iff $\lambda_i \in \R$ with $\sum_{i=1}^{m+1} \lambda_i = 0$ and $\sum_{i=1}^{m+1} \lambda_i v_i = 0$ implies $\lambda_i = 0 \ (\forall i=1,\ldots,m+1)$. 
It is an easy exercise to show that the followings are all equivalent: (i) $\{v_i\}_{i=1}^{m+1}$ is an affinely independent set; 
(ii) For any $i_0$, $\{v_i - v_{i_0 } \}_{i\neq i_0}$ is a linearly independent set; (iii) Convex decomposition w.r.t. $\{v_i\}_{i=1}^{m+1}$ is unique. 
(Namely, if $\sum_{i=1}^{m+1} p_i v_i = \sum_{i=1}^{m+1} q_i v_i $ with probability distributions $(p_i)_i$ and $(q_i)_i$, then $p_i = q_i \ ( \forall i=1,\ldots,m+1)$).}.  
  
Any classical state ${\bm p} = (p_1,\ldots,p_c) \in \SA_{cl}$ has the unique decomposition into pure states: 
${\bm p} = \sum_{j=1}^c p_j {\bm p}^{(j)}$, and it is convenient to represent an effect $e$ by $c$ dimensional vector $(x_1,\ldots,x_c) \in [0,1]^c$ by $x_i := e({\bm p}^{(j)})$ with $e({\bm p}) = \sum_j p_j x_j$. 
We write $e \simeq (x_1,\ldots,x_c)$ in this case. 
Note that $u \simeq (1,\ldots,1)$ and $0 \simeq (0,\ldots,0)$. 
It is easy to see that $\E(\SA_{cl}) =: \E_{cl} \simeq [0,1]^c$, so there are $2^c$ decision effects (i.e., extreme points of $\E_{cl}$): $e^{i_1 i_2 \cdots i_c} \simeq (i_1,i_2,\ldots,i_c) $ where $i_1,\ldots,i_c = 0,1$. 
Any measurement $M=(e_i)_{i=1}^n$ is represented by $n$ vectors $e_i \simeq (x^{(i)}_1,\ldots,x^{(i)}_c) \in [0,1]^{c}$ such that $\sum_j x^{(i)}_j = 1$. 
Typical measurement is given by that of random variable $f: \Omega_c \to \{m_1,\ldots,m_n\}$ where $\Pr\{f = m_i | {\bm p}\}:= \sum_{\scriptsize j; m_i = f(j)} p_j = \sum_{i=1}^c \delta_{m_i f(j)} p_j$. 
Since $(\delta_{m_i f(j)})_{j=1}^c$ is one of decision effects, a measurement of random variable corresponds to a measurement with respect to decision effects.

Note that any $c$-simplex is affine isomorphic to the $c$-dimensional standard simplex. 
Therefore, any GPT with simplex-state space is a classical system.

[Example 2] (Finite Quantum Mechanics) A state of a finite QM is represented by a density operator $\rho$ (a positive operator with unit trace, denoted as $\rho \ge 0$ and $\tr \rho = 1$) on a $c$ dimensional Hilbert space $\HA_c$. 
The state space is thus given by $\SA_q:= \{\rho \in \LA(\HA_c)\ | \ \rho \ge 0, \tr \rho = 1\} \subset \LA(\HA_c)_h$, where $\LA(\HA_c)$ $(\LA(\HA_c)_h)$ is the set of all the linear (Hermitian) operators on $\HA_c$.  
Pure states correspond to the one dimensional projection operators, i.e.,  $\rho=\ketbra{\psi}{\psi}$ with unit vector $\psi \in \HA_c$. 
A quantum effect $e$ is usually represented by a positive operator $E \in \LA(\HA_c)$ such that $e(\rho) = \tr (E \rho)$ given by $E:= \sum_{i,j=1}^c \tilde{e}(\ketbra{\psi_i}{\psi_j}) \ketbra{\psi_j}{\psi_i}$ with an arbitrary orthonormal basis $\{\psi_i\}_{i=1}^c$ of $\HA_c$, where $\tilde{e} $ is a linear extension of $e$ to $\LA(\HA_c)$ (See Appendix \ref{app:ExAff}).  
Note that the unit and zero effects $u,0$ correspond to the unit and zero operators $\I $ and $0$, respectively.  
It is easy to see that $\E_q:= \E(\SA_q) \simeq \LA(\HA_c)_{povm} := \{E \in \LA(\HA_c) | 0 \le E \le \I \}$, the element of which is called a POVM (positive operator valued measure) element. 
Thus, any measurement $M = (e_i)_{i=1}^n $ is represented by $n$-tuple of POVM elements $ e_i \simeq E_i \in \LA(\HA_c)_{povm} $ such that $\sum_{i=1}^n E_i = \I $, which is called a (discrete) POVM.   
The decision effects correspond to projection operators. 

The most elementary quantum system is a qubit system with $2$ dimensional Hilbert space $\HA_2$. 
For a qubit system, the Bloch vector ${\bm b}=(b_1,b_2,b_3) \in \R^3$, defined by $\SA_{q} \ni \rho \mapsto b_i= \tr(\sigma_i \rho)$ $(i=1,2,3)$, gives a useful state representation. (Here, $\sigma_i \ (i=1,2,3)$ are Pauli spin matrices). 
Remind that the state space is then given by a unit ball $B:=\{{\bm b} \in \R^3 | \sum_i b_i^2 = 1\} \subset \R^3$ \cite{ref:NC}. 

Notice again that both classical and quantum theories satisfy [P4], i.e., any mathematically well-defined measurement corresponds to a feasible measurement by admitting the so-called indirect measurement 
\footnote{In QM, we usually admit, especially in finite systems, that any Hermitian operator (called an observable) and unitary operation correspond to feasible measurement and dynamics, respectively. 
Then, given any POVM $(E_i)_{i=1}^n$ in quantum system, there exists an ancilla system described by Hilbert space $\HA_n$, an initial state $\sigma$ on $\HA_n$, and unitary operator $U$ on $\HA_{tot}:= \HA_d\otimes \HA_n$, and a meter observable $M = \sum_i m_i P_i $ on $\HA_n$ such that $\tr(E_i \rho) = \tr_{tot}( U(\rho \otimes \sigma)U^\dagger \I \otimes P_i)$. }.



\section{Additional Principles}\label{sec:AP}

In the previous section, GPTs are explained as operationally well-defined probability models based on natural principles [P1]-[P4]. 
In this section, we further consider additional two principles of [P5] {\it physical equivalence of pure states} and [P6] {\it decomposability with distinguishable pure states}. 

\subsection{Physical Equivalence of Pure States}\label{subsec:PE}

Intuitively, it seems natural to assume that all the pure states are physically equivalent, i.e., there are no physically exceptional pure states in nature due to its symmetry. 
This would be acceptable if one reminds both classical and quantum cases, where in both cases there are no structural differences in the set of pure states. 
(For instance, consider a one dimensional classical particle, where the set of pure states is the phase space $\R^2$ of the position and the momentum. 
Obviously, there are no exceptional points in $\R^2$. 
Also in quantum systems, there are no exceptional vectors (pure states) in Hilbert space.)  
As this argument yet resorts to a rough intuition, we consider here what we exactly mean by the physically equivalence of states with an operational view point of measurements.   

First of all, we would like to express that states $s_1,s_2 \in \SA$ are physically equivalent if there are no different physical structures on $s_1$ and $s_2$ through measurements. 
This would be naturally characterized by saying the following: 
For any measurement $E = (e_i)_{i=1}^n$, there uniquely exists the corresponding measurement $F=(f_i )_{i=1}^n$ such that the probability distributions on $s_1$ by $E$ and $s_2$ by $F$ are the same: $e_i(s_1) = f_i(s_2)$; 
Next, this correspondence of the measurements should preserve the affine structures for a convex combination of measurements. 
Indeed, one can perform a convex combination of measurements by a probabilistic mixture of each measurement. 
Therefore, the corresponding measurement of a convex combination of $E$ and $E'$ should also be performed by the probabilistic mixture of $F$ and $F'$ (the corresponding measurements of $E$ and $E'$) with the same weight. 
Finally, the correspondent of the trivial measurement $(u)$ (i.e., $1$-valued measurement) should be again the trivial measurement $(u)$. 
From these considerations (in particular for yes-no measurements), we can give the formal definition of state-equivalence as follows:
\begin{Def}\label{def:PE} (Physical Equivalence of States)
We say that state $s_1$ is physically equivalent to state $s_2$ if there exists a unit-preserving affine bijection $\Phi:\E \to \E$ such that 
$e(s_1) = \Phi(e)(s_2)$ for any $e \in \E$. 
\end{Def}
We denote by $s_1 \simeq s_2$ iff $s_1$ is physically equivalent to $s_2$. 
It is easy to see that $\simeq$ is an equivalence relation: (i) $s \simeq s$, (ii) $s_1 \simeq s_2 \Rightarrow s_2 \simeq s_1$ and (iii) $s_1 \simeq s_2, s_2 \simeq s_3 \Rightarrow s_1 \simeq s_3$ for any $s,s_1,s_2,s_3 \in \SA$. (Use identity map on $\E$, an inverse map and composition of maps on $\E$ as the above map $\Phi$):  
\begin{Prop}
Physical equivalence $\simeq$ is an equivalence relation.  
\end{Prop}
Based on this operationally motivated definition, we consider the following principle: 

[P5] (Physical Equivalence of Pure States) {\it Any pure states $s_1$ and $s_2 $ are physically equivalent. }  

The following result gives an equivalent characterization of the physical equivalence in terms of the structure of state space: 
\begin{Thm}\label{thm:PESym}
States $s_1$ and $s_2$ are physically equivalent iff there exists an affine bijection $\Psi :\SA \to \SA$ such that $s_1 = \Psi(s_2)$. 
\end{Thm}

To prove this theorem, we start from the following fact on effects: 
The set of effect $\E= \E(\SA)$ has a natural order relation defined by 
$$
e \ge f \Leftrightarrow e(s) \ge f(s) \ \forall s \in \SA
$$
for effects $e,f \in \E$. 
Note that $u$ and $0$ are maximum and minimum effects. 

\begin{Lem}\label{lem:bPhi}
Let $\Phi:\mathcal{E} \to \mathcal{E}$ be an affine bijection.
Then the following three conditions are equivalent:

$\mathrm{1}$. $\Phi(0) = 0$ 

$\mathrm{2}$. If $e,f \in \mathcal{E}$ and $e \leq f$, then $\Phi(e) \leq \Phi(f)$ 

$\mathrm{3}$. $\Phi(u) = u$ 
\end{Lem}
(The proof is given in Appendix \ref{sec:A}. This lemma is used just for justifying that we have assumed in Definition \ref{def:PE} only the latter one of the two conditions $\Phi(0) = 0$ and $\Phi(u) = u$.)
 
Now we introduce the dual maps of affine bijections on $\E$ and $\SA$, respectively. 

First, let $\Phi: \E \to \E$ be an affine bijection on $\E$ such that $\Phi(u) = u$ (and thus $\Phi(0) = 0$ from Lemma \ref{lem:bPhi}). 
Then, the dual map $\Phi^\ast:\SA \to \SA$ on $\SA$ is defined by 
\begin{equation}\label{eq:dual}
\Phi^\ast(s) (e) := s(\Phi(e)) \ (\forall s \in \SA, e \in \E). 
\end{equation}
(See Eq.~\eqref{eq:se=es} for the notation.)
We notice that Eq.~\eqref{eq:dual} for fixed $s \in \SA$ defines an affine functional on $\E$ satisfying conditions in Theorem \ref{thm:AF}. 
To see this, let $\Lambda:= \Phi^\ast(s)$. 
For any $\lambda \in [0,1],e_1,e_2 \in \E$, we have $\Lambda(\lambda e_1 + (1-\lambda) e_2) = s(\Phi(\lambda e_1 + (1-\lambda) e_2) ) = s(\lambda  \Phi(e_1) + (1-\lambda) 	\Phi(e_2) ) =  \lambda s ( \Phi(e_1)) + (1-\lambda) s( \Phi(e_2)) = \lambda \Lambda(e_1) + (1-\lambda) \Lambda(e_2)$, and thus $\Lambda$ is an affine functional on $\E$. 
Since $s(e) \in [0,1]$ for any $e \in \E$, the range of $\Lambda$ is in $[0,1]$. 
Finally, $\Lambda(u) = s(\Phi(u)) = s(u) = 1$ and $\Lambda(0) = s(\Phi(0)) = s(0) = 0$. 
Thus $\Phi^\ast$ is a well-defined map on $\SA$ from Theorem \ref{thm:AF}. 

Notice that, for any $\lambda \in [0,1], s_1,s_2 \in \SA$, one has $
\Phi^\ast(\lambda s_1 + (1-\lambda) s_2) (e) = (\lambda s_1 + (1-\lambda) s_2) \Phi(e) = (\lambda \Phi^\ast(s_1)  + (1-\lambda) \Phi^\ast(s_2))(e) \ (\forall e \in \E)$, 
hence $\Phi^\ast$ is affine on $\SA$. 
Let $\Phi^\ast(s_1) = \Phi^\ast(s_2)$ for some states $s_1,s_2 \in \SA$. 
Then, since $f:= \Phi^{-1}(e) \in \E$ for any $e \in \E$, we have 
$s_1(e)= s_1(\Phi(\Phi^{-1}(e))) = \Phi^\ast(s_1)(f) = \Phi^\ast(s_2)(f) = 
s_2(\Phi(\Phi^{-1}(e))) = s_2(e)$. 
Therefore, from the separation hypothesis for states, we have $s_1 = s_2$, i.e., 
$\Phi^\ast$ is injective (one-to-one).  
Notice that $\Phi^{-1}$ is also an affine bijection on $\E$ such that $\Phi^{-1}(u) = u $ and $\Phi^{-1}(0) = 0 $. 
For an arbitrary $s \in \SA$, define $t:= (\Phi^{-1})^\ast (s) \in \SA$, and it follows that for any $e \in \E$
$$
\Phi^\ast(t)(e) = t (\Phi(e)) = s(\Phi^{-1}(\Phi(e))) = s(e), 
$$
and again from the separation hypothesis for states, we have $s = \Phi^\ast(t)$. Therefore, $\Phi^\ast$ is surjective, hence an affine bijection on $\SA$. 

On the other hand, the dual map of the affine bijection $\Psi :\SA \to \SA$ is easily defined by 
\begin{equation}\label{eq:DPsi}
\Psi^\ast(e)(s):= e(\Psi(s)) \ (\forall e\in \E, s \in \SA). 
\end{equation}
Since the above defined $\Psi^\ast(e)$ is obviously an affine functional on $\SA$ to $[0,1]$, $\Psi^\ast(e) $ is an effect. 
It is easy to see that $\Psi^\ast$ is bijective on $\E$ and satisfies $\Psi^\ast(u)=u$ (and $\Psi^\ast(0)=0$). 

Finally, it is also easy to see that both dual maps on $\E$ and $\SA$ satisfy $(\Phi^\ast)^\ast = \Phi$ and $(\Psi^\ast)^\ast = \Psi$ from the definitions. 
To sum up, we have obtained
\begin{Prop}\label{prop:duals}
(i) For any affine bijection $\Phi: \E \to \E$ such that $\Phi(u) = u $ (and thus $ \Phi(0) = 0$), the dual map $\Phi^\ast:\SA \to \SA$ is well-defined by Eq.~\eqref{eq:dual} and is affine bijective. 
(ii) For any affine bijection $\Psi:\SA \to \SA$, the dual map $\Psi^\ast$ is well-defined by Eq.~\eqref{eq:DPsi} and is affine bijective satisfying $\Psi^\ast(u)=u$ (and thus $\Psi^\ast(0)=0$). 
(iii) For both bijections, $(\Phi^\ast)^\ast = \Phi$ and $(\Psi^\ast)^\ast = \Psi$.   
\end{Prop}

Now we can give the proof of Theorem \ref{thm:PESym}: 

\noindent {\bf [Proof of Theorem \ref{thm:PESym}]} Let $s_1,s_2$ be physically equivalent, i.e., there exists a bijective affine $\Phi:\E\to \E$ such that $e(s_1) = \Phi(e)(s_2)$ for any $e \in \E$ and $\Phi(u) = u$ (and thus $\Phi(0) =0$ from Lemma \ref{lem:bPhi}). 
From Proposition \ref{prop:duals} (i), $\Psi:= \Phi^\ast$ gives an affine bijection on $\SA$. 
From Eq.~\eqref{eq:dual}, we have $\Psi(s_2)(e) = s_2(\Phi(e)) = \Phi(e)(s_2) = e(s_1) = s_1(e)$ for any $e \in \E$, and thus $s_1 = \Psi(s_2)$.  

Conversely, assume that there exists an affine bijection $\Psi:\SA \to \SA$ such that $s_1 = \Psi(s_2)$. 
Then, from Proposition \ref{prop:duals} (ii), the dual map $\Phi:= \Psi^\ast $ is an affine bijection on $\E$ satisfying that $\Phi(e)(s_2) = \Psi^\ast(e)(s_2) = e(\Psi(s_2)) = e(s_1)$ and $\Phi(u) = u$. 
Therefore, states $s_1,s_2$ are physically equivalent.  
\hfill $\blacksquare$

From Theorem \ref{thm:PESym}, [P5] is equivalently rephrased as follows: 

[P5']{\it For any given pure states $s_1,s_2\in \SA$, there exists a bijective affine map $\Psi$ on $\SA$ such that $s_2 = \Psi(s_1)$. } 

In \cite{ref:KNI}, we introduced the principle of equality of pure states intuitively meaning [P5] but have used [P5'] as its formal definition. 
Theorem \ref{thm:PESym} establishes its operational meaning as an equivalence of pure states. 
Some authors \cite{ref:Hardy,ref:DakicBrukner,ref:Masanes} also have used [P5'] meaning that there are reversible transformations (dynamics) which connects any pure states. 
Thus, we have shown the equivalence of two superficially different physical principles. 

On the other hand, from the mathematical and physical interests, Davies \cite{ref:Davies} have investigated the following condition: 

[P5''] The group of affine bijections on $\SA$ acts transitively on the set of extreme points of $\SA$. 

It turns out that [P5], [P5'], and [P5''] are all equivalent: 
To see this, let $\A_b(\SA,\SA)$ be the set of affine bijections on $\SA$. 
One can show that $\A_b(\SA,\SA)$ forms a compact group which acts continuously on $\SA$ \cite{ref:Davies}.  
 Note that an affine bijection on $\SA$ maps a pure state to a pure state \footnote{Let $\Psi:\SA \to \SA$ be an affine bijection and let $s \in \SA$ be a pure state. 
Let $\Psi(s) = p s_1 + (1-p)s_2$ with some $t_1,t_2 \in \SA$ and $p \in (0,1)$. 
Applying $\Psi^{-1}$, we have $s = p \Psi^{-1}(s_1) + (1-p) \Psi^{-1}(s_2)$ and thus $s = \Psi^{-1}(s_1) = \Psi^{-1}(s_2)$, as $s$ is a pure state. 
Therefore, we have $\Psi(s) = s_1 = s_2$, and $\Psi(s)$ is a pure state. }. 
Therefore, it is straightforward that [P5'] holds iff affine bijections on $\SA$ acts transitively on $\SA_{pure}$, i.e., $\SA_{pure} = \{\Psi (s) \ | \ \Psi \in \A_b(\SA,\SA) \}$ with arbitrary reference pure state $s \in \SA_{pure}$.

In the following, we prefer to interpret the above equivalent conditions as a physical equivalence among any pure states [P5]. 
Indeed, as is the case for any basic physical theories so far, a symmetry is one of the key factor to understand our world, and [P5] enables us to interpret, without appealing to dynamics, that there exists an operationally motivated symmetry in the set of pure states. 
Of course, this is a matter of taste, and the following argument follows for those who prefer [P5'] or [P5'']. 
However, following Davies, we call GPTs satisfying [P5] (or equivalently [P5'] or [P5'']) {\it symmetric} GPTs.


\subsubsection{Examples of Symmetric GPTs}\label{sec:ESGPT}

Both classical theory and QM are symmetric GPTs. 
Indeed, for any pure states ${\bm p}^{(\mu)},{\bm p}^{(\nu)} \in \SA_{cl}$ with some $\mu,\nu = 1,\ldots,d$, let $\Psi$ be a map on $\SA_{cl}$ defined by 
$[\Psi({\bm p})]_\mu = p_\nu, [\Psi({\bm p})]_\nu = p_\mu$ and $[\Psi({\bm p})]_i = p_i \ (i \neq \mu,\nu)$. 
Obviously $\Psi$ is an affine bijection on $\SA_{cl}$ satisfying $\Psi({\bm p}^{{\scriptsize \mu}}) = {\bm p}^{\nu}$. 
In quantum system, for any pure states $\rho_1:= \ketbra{\psi_1}{\psi_1},\rho_2:= \ketbra{\psi_2}{\psi_2} \in \SA_{q}$, one can easily find a unitary operator $U:\HA_d \to \HA_d$ such that $U \psi_2 = \psi_1$ \footnote{For instance, let $\{\phi_i\}_{i=1}^d$ and $\{\phi'_j\}_{j=1}^d$ be orthonormal bases of $\HA_d$ such that $\phi_1 = \psi_1$ and $\phi'_1 = \psi_2$. Then, use a unitary operator $U:= \sum_{i=1}^d \ketbra{\phi_i}{\phi'_i}$}.  
Then it is easy to see that $\Psi(\rho):= U \rho U^\dagger$ is an affine bijection on $\SA_q$ satisfying $\rho_1 = \Psi(\rho_2)$.  

Another simple example of symmetric GPTs are cuboid systems \cite{ref:KMI}: 

\noindent [Example 3] (Cuboid Systems) We call GPTs cuboid systems if the state spaces are represented by hypercubes: 
$\SA_{cube}:= \{{\bm x} \in \R^d \ | \ 0 \le x_i \le 1 \ ( i=1,\ldots,d ) \}$. 
There are $2^d$ pure states (vertices): ${\bm x}^{(i_1,i_2,\ldots,i_d)}:= (i_1,i_2,\ldots,i_d)  \ (i_1,\ldots,i_d = 0,1)$.

It is easy to see that the cuboid system $\SA_{cube}$ is also symmetric. 
In general, any isogonal figure is symmetric from the definition: 

\noindent [Example 4] (Systems with Isogonal Figures)
A polytope in Euclidean space is said to be isogonal iff for any vertices there exists an isometrical map which surjectively connects them \cite{ref:IsogonalPolytope}. 
(Notice that any isometrical map on Enclidean space is automatically affine.) 
Therefore, any GPTs with state space being isogonal figure are symmetric.

\subsubsection{Invariant States}

In this section, we discuss an invariant state (the maximal mixed state) in GPTs. 

In general GPTs, the existence of an invariant state with respect to all the bijective affine maps on $\SA$ can be shown based on Ryll-Rardzewski fixed point theorem.  
\begin{Thm}\label{thm:ExIS}
For any GPT, there exists an invariant state $s_{inv} \in \SA$ with respect to all the bijective affine maps on $\SA$.  
\end{Thm}
{\bf [Proof]} \ 
We refer to \cite{ref:RN,ref:RN2} for Ryll-Rardzewski fixed point theorem. 
Using the norm in Appendix \ref{app:Rep2}, the underlying vector space $V$ of $\SA$ is a Banach space w.r.t. the sup-norm and $\SA$ is a compact convex subset of $V$. 
Since we assumed dim $V < \infty$, all topologies in the Ryll-Nardzewski theorem are equivalent to the norm topology. 
Let $G=A_b(\SA,\SA)$ be the set of all the bijective affine maps on $\SA$. 
Then, it is easy to show that $G$ is a semigroup of (weakly) continuous affine maps of $\SA$. 	
In particular, $G$ is noncontractive on $\SA$, since the norm $||\cdot ||$ satisfies $\inf_{\Phi \in G } ||\Phi (s_1) - \Phi (s_2)|| = ||s_1 - s_2 || > 0 \ (\forall s_1 \neq s_2 \in \SA)$ from Lemma \ref{lem:InvNorm} in Appendix \ref{app:Rep2}. 
Applying the Ryll-Nardzewski theorem, the assertion of the theorem holds. 
\hfill $\blacksquare$

In general GPT, an invariant state is not necessarily unique. 
(For instance, consider a GPT with the state space being a circular sector $\SA = \{(r \cos \theta, r \sin \theta) | 0 \le r \le 1, 0 \le \theta \le \frac{\pi}{2}\}$. 
Then, there are only two affine bijections on $\SA$: the identity map and the reflection map with respect to the $45$ degree axis. Then, all the states on the axis are invariant under these maps.) 
However, in a symmetric GPT, the unique invariant state is determined.  
\begin{Thm}
In a symmetric GPT, an invariant state is unique. 
\end{Thm}
{\bf [Proof]} \ 
Assume that there exist distinct invariant states $s_1,s_2$ on $\SA$. 
Then, any state of the form $p s_1 + (1-p) s_2 \in \SA$ for any $p \in \R$ is invariant under any bijective affine $\Psi: \SA \to \SA$ (See Lemma \ref{lem:rest} in Appendix \ref{app:ExAff}).  
In particular, there exists such state $s_3$ on a boundary of $\SA$. 
Let $H \subset V$ be the supporting hyperplane of $\SA$ on $s_3$ and let $s_3 = \sum_{i=0}^{n} p_i t_i \ (p_i >0) \ (n \in \N)$ be a convex decomposition of $s_3$ with respect to pure states $\{t_i\}$. 
Then, since $s_3 \in H$ and $H$ is a supporting hyperplane of $\SA$, we have $t_i \in H \cap \SA$ for every $t_i$.  
Notice that there exists a pure state $s_0 \in \SA$ not in $H$. (Otherwise, $\SA \subset H$.)
Since $\SA$ is symmetric, there exists an affine bijective map $\Psi$ which connects $t_0$ and $s_0$: $s_0 = f(t_0)$. 
However, we have $\sum_{i=0}^{n} p_i t_i = s_3 = f(s_3) = \sum_{i=0}^{n} p_i f(t_i) = p_0 s_0 + p_1 f(t_1) \cdots$, and this is contradictory.  
\hfill $\blacksquare$

In the following, $s_M$ denotes an invariant state, which is uniquely determined for symmetric GPTs.  
For a GPT with polytope state space, where the set of pure states is  $\{s_1,\ldots,s_n\}$, we have 
\begin{equation}
s_M = \frac{1}{n}\sum_{i=1}^n s_i. 
\end{equation}
Indeed, as any bijective affine $\Phi$ on $\SA$ maps any pure state to a pure state, 
$\Phi$ just permutes finite numbers of pure states, and we have 
$\Phi(s_M) = \frac{1}{n}\sum_{i=1}^n \Phi(s_i) = s_M$.  
In particular, in classical system, $s_M$ is the uniform distribution on $\Omega$: $s_M = (1/d,\ldots,1/d) $. 
For cuboid systems, we have $s_M = (1/2,1/2,\ldots,1/2)$, i.e., the center of $\SA_{cube}$.  

In the case of finite quantum systems, notice that bijective affine map on $\SA_q$ is 
unitary map or anti-unitary map on the density operators \cite{ref:Kadison}. 
Therefore, the maximally mixed state 
$$
s_M = \frac{1}{d} \I
$$
is the unique invariant state. 

In general, the construction \cite{ref:Davies} of $s_M$ is given by the two-sided Haar measure \cite{ref:UOH}: 
Using the two-sided Haar measure on the set $G$ of all the affine bijection on $\SA$, we have 
\begin{equation}\label{ref:MMS}
s_M = \int_G \Psi(s_0) d\mu(\Psi),
\end{equation}
where $s_0 \in \SA $ is an arbitrary reference state (See also \cite{ref:NK} for the mathematical details). 
Indeed, the unique existence of the invariant states of Symmetric GPTs is first shown by Davies \cite{ref:Davies} by means of the Haar measure. 

Note that an operationally natural entropy in general GPTs is defined by 
\begin{equation}\label{eq:entropy}
S(s):= \inf_{M=(m_j)_j\in M_{ind}} H(m_j(s)),
\end{equation}
where $H(p_j) := -\sum_j p_j \log_2 p_j $ is the Shannon entropy and the infimum is taken over the set $M_{ind}$ of all the indecomposable measurements \cite{ref:KNI,ref:Entropies}. 
$S$ is concave on $\SA$, generalizing Shannon and von Neumann entropies in classical and quantum systems, respectively. 
In a symmetric GPT, $S$ takes minimum value $0$ on pure states, and thus $S$ provides a proper measure of mixedness \cite{ref:KNI}. 
Therefore, the following proposition implies that we may call $s_M$ the maximal mixed state in general (symmetric) GPT.
\begin{Prop}
In any GPT, the entropy $S$ takes the maximal value on an invariant state. 
\end{Prop}
{\bf [Proof]} \ 
Notice that for a bijective affine $\Psi$ on $\SA$ and indecomposable measurement $M=(m_j)_j$, $(m_j \circ \Psi)_j$ is an indecomposable measurement \cite{ref:KNI}. 
Thus, we have $S(\Psi(s)) = S(s) $ for any state $s$. 
Then, from the concavity of $S$, for any state $s$, the invariant state $s_M$ obtained by (\ref{ref:MMS}) from the reference state $s_0 := s$ satisfies
$$
S(s_M) \ge \int_G S(\Psi(s)) d\mu (\Psi) = \int_G S(s) d\mu (\Psi)  = S(s). 
$$
\hfill $\blacksquare$


The next result shows a global picture of a state space $\SA$ in symmetric GPT. 
Let $D$ be a distance function which have the monotonicity properties under affine bijection on $\SA$, i.e., for any states $s_1,s_2 \in \SA$ and any affine bijection $\Psi : \SA \to \SA$, $D(s_1,s_2) \ge D(\Psi(s_1),\Psi(s_2))$. 
The typical example is given by the Kolmogorov distance (See Appendix \ref{app:Rep2}). 
\begin{Prop}\label{prop:INVDIS}
In symmetric GPTs, for any distance function $D$ on $\SA$ with the monotonicity property, 
the distances between $s_M$ and arbitrary pure states are constant. 
Namely, for any pure states $s_1,s_2 $, 
\begin{equation}\label{eq:DConst}
D(s_1,s_M) = D(s_2,s_M)
\end{equation}
\end{Prop}
{\bf [Proof]} \ From the symmetry of $\SA$, there exists an affine bijection $\Psi$ on $\SA$ such that $s_1 = \Psi(s_2)$. 
From the monotonicity of $D$ and the invariance of $s_M$, we have 
$$
D(s_1,s_M) = D(\Psi(s_2),\Psi(s_M)) \le D(s_2,s_M).
$$
The opposite inequality follows by a symmetric argument. \hfill $\blacksquare$

Geometrically speaking, all the pure states of symmetric GPTs lie on the sphere with respect to the distance with the monotonicity property, e.g. Kolmogorov distance.
In the case of Kolmogorov distance \eqref{eq:KolDis}, physical meaning of \eqref{eq:DConst} is that the optimal success probability to distinguish between the invariant state $s_M$ and any pure state $s$ is constant on $s$ (See Eq.~\eqref{eq:KolDis} in Appendix \ref{app:Rep2}).  

\subsubsection{Classification of Symmetric GPTs}

Now we proceed to the classification theorem of Symmetric GPTs. 
First, remind that without loss of generality, $\SA$ is a compact subset of $\R^d$, and by the parallel shift, we can assume the invariant state $s_M$ of $G$ to be the origin of $\R^d$.   

Using the Haar measure on the set of $G$ of all the affine bijection on $\SA$, one can define a $G$-invariant inner product by 
$\langle x,y\rangle_G := \int_G \langle \Psi x, \Psi y\rangle d \mu(\Psi)$ where the right-hand side is the Haar integral and $\langle x,y\rangle$ denotes the Euclidean inner product of $x,y \in \R^d$. 
Then, using an orthonormal basis w.r.t. this $G$-invariant basis, we get to a new state-representation in $\R^d$ such that $\Psi \in G$ acts isometrically on states. 
Remind that, in Sec.~\ref{sec:ESGPT}(Example 4), we have just illustrated symmetric GPTs with state spaces being isogonal figures, but now it earns the general symmetric GPTs if the numbers of pure states are finite:  
\begin{Prop}\label{Prop:IF}
The state space $\SA$ of a symmetric GPT is either one of isogonal figures or with infinite pure states. 
\end{Prop}
In \cite{ref:NK}, we further obtain the following classification theorems of symmetric GPTs in $2$ and $3$ dimensional cases: 
\begin{Thm}\label{thm:CSym2}
The state space $\SA$ of symmetric GPTs with $\dim(\mathrm{Aff}\SA) = 2$ is either an isogonal figure or a unit disk. 
\end{Thm}
\begin{Thm}\label{thm:CSym3}
The state space $\SA$ of symmetric GPTs with $\dim(\mathrm{Aff} \SA) = 3$ is any one of an isogonal figure, a circular cylinder, or a unit disk (ball). 
\end{Thm}

\subsection{Decomposability of states with distinguishable pure states} 

In this section, we consider another principle of the decomposability of states with respect to distinguishable pure states. 

Remind that states $\{s_1,\ldots,s_n\}$ are said to be distinguishable if they can be distinguishable with probability $1$ in a single shot measurement;  
namely if there exists a measurement $M = (e_i)_{i=1}^n$ such that $e_i(s_j) = \delta_{ij}$ for any $i,j = 1,\ldots,n$. 
The maximum number of distinguishable states is an important parameter to characterize GPTs, and is denoted as $c$.  
In classical systems, $c$ is the number (cardinality) of the sample space $\Omega_c$, while in quantum systems it is the dimension of the corresponding Hilbert space $\HA_c$. 
In general GPT, $c$ has the following upper bound:    
\begin{Prop}\label{Prop:dimc}
In GPT with state space $\SA$, we have 
\begin{equation}\label{eq:c<dimS+1}
c \le \dim(\SA) + 1. 
\end{equation}
The equality holds iff $\SA$ is a simplex.  
\end{Prop}

To see this, observe the following lemma: 
\begin{Lem}\label{lem:dissi}
A distinguishable set of states $\{s_1,\ldots,s_n\}$ is affinely independent. Therefore, $\mathrm{Conv} \{s_i\}$ is an $(n-1)$ dimensional simplex. 
\end{Lem}
{\bf [Proof]} \ Let $\sum_{i=1}^n p_i s_i = \sum_{i=1}^n q_i s_i $ be convex combinations of $\{s_i\}_i$ w.r.t. probability distributions $(p_i)_i$ and $(q_i)_i$. 
Then, applying a measurement $M = (e_i)_i$ which distinguishes $\{s_i\}_i$, we get $p_i = q_i$, and thus $\{s_i\}_i$ is affinely independent (See [39]). 
\hfill $\blacksquare$ 

\noindent {\bf [Proof of Proposition \ref{Prop:dimc}]} \ 
Let $\{s_1,\ldots,s_c\}$ be a set of distinguishable states. 
Then, from Lemma \ref{lem:dissi}, $\mathrm{Conv} \{s_i\}$ is a $(c-1)$ dimensional simplex. 
As $\mathrm{Conv} \{s_i\} \subset \SA$, we have $c-1 \le \dim (\SA)$. 

Let the equality of Eq.~\eqref{eq:c<dimS+1} holds. 
Let $d:= \dim (\SA)$ and let $\{s_1,\ldots,s_{d+1}\}$ be distinguishable states with a measurement $M=(m_j)_{j=1}^{d+1}$ which distinguishes them. 
Note that, from Lemma \ref{lem:dissi}, Conv $\{s_i\}_{i=1}^{d+1}$ is a $d$-dimensional simplex. 
We show that $\SA = \mathrm{Conv} \{s_i\}_{i=1}^{d+1}$. 
To see this, notice that the dimension of Aff $\{s_i\}_{i=1}^{d+1}$ is $d$ 
and $\{s_i\}_{i=1}^{d+1} \subset \SA$, and thus Aff $\{s_i\}_{i=1}^{d+1}= \mathrm{Aff} \SA$. 
Therefore, any $s \in \SA$ can be written as $s = \sum_{i=1}^{d+1} \lambda_i s_i $ with some $\lambda_i \in \R $ satisfying $\sum_i \lambda_i = 1$. 
Applying $m_i$ to $s$ (See Lemma \ref{lem:rest} in Appendix \ref{app:ExAff}), we obtain $\lambda_i = m_i(s) \ge 0$, 
and thus $s \in \mathrm{Conv} \{s_i\}_{i=1}^{d+1}$. 
Therefore, we have $\SA \subset \mathrm{Conv} \{s_i\}_{i=1}^{d+1}$. 
As $s_i \in \SA$, we obtain $\SA = \mathrm{Conv} \{s_i\}_{i=1}^{d+1}$. 
Namely, $\SA$ is $d$ dimensional simplex. \hfill $\blacksquare$

In both classical and quantum systems, any state is written as a convex combination of distinguishable pure states. (Note that any classical state ${\bm p} = (p_1,\ldots,p_d)$ has the decomposition ${\bm p} = \sum_{\mu=1}^d p_\mu {\bm p}^{\mu} $, 
while any quantum state $\rho$ has a decomposition $\rho = \sum_{i=1}^c p_i \ketbra{\psi_i}{\psi_i}$ as an eigenvalue decomposition of $\rho$.) 
This property can be physically phrased as following physical principle:   

[P6] (Decomposability with Distinguishable Pure States) {\it Any state can be prepared as a probabilistic mixture of distinguishable pure states}. 

It is interesting to consider GPTs satisfying this principle, which include both classical and quantum systems. 
We notice that this property is derived from their axioms in \cite{ref:CDP}.

The following result provides the general classification of such GPTs:  
\begin{Thm}\label{thm:P6}
GPTs satisfying [P6] is either classical system or systems with infinite pure states.  
\end{Thm}

To prove this, we use the following lemma: 
\begin{Lem}\label{lem:PolyD+1}
If $\SA$ is a $d$ dimensional polytope satisfying the principle [P6], then there exist $d+1$ distinguishable states. 
\end{Lem}

\noindent {\bf [Proof]} \ The case $d = 0$ is trivial.
Let $d \ge 1$. 
Let $\mathcal{F}$ be the family of all subsets of $\SA$ such that it is the convex hull of at most $d$ pure states of $\SA$.
Since $\SA$ is a polytope, the number of pure states is finite, therefore the size of $\mathcal{F}$ is also finite. 
Now each subset of $\SA$ in $\mathcal{F}$ is at most ($d-1$)-dimensional, therefore its $d$-dimensional volume is $0$.
Hence the union $\bigcup \mathcal{F}$ of all (finitely many) members of $\mathcal{F}$ also has $d$-dimensional volume $0$, 
while $\SA$ has positive $d$-dimensional volume.
This implies that $\SA$ contains a state $s$ that does not belong to $\bigcup \mathcal{F}$.
By the principle of [P6], $s$ is a convex combination of distinguishable pure states $s_1,\dots,s_k$, and now $k > d$ by the definition of $\mathcal{F}$ and $s$.
Hence the claim holds. 
\hfill $\blacksquare$

\noindent {\bf [Proof of Theorem \ref{thm:P6}]} \ 
Let the number of pure states be finite, i.e., $\SA$ be a polytope. 
From Lemma \ref{lem:PolyD+1}, there exist $d+1$ distinguishable states. 
Therefore, the equality of Eq.~\eqref{eq:c<dimS+1} holds and thus from Proposition \ref{Prop:dimc}, $\SA$ is simplex, i.e., a state space with a classical system. 
\hfill $\blacksquare$

\subsection{Classification of GPTs with [P5] and [P6]}

Combining the classifications of GPTs with [P5] and [P6], i.e., Theorem \ref{thm:CSym2}, Theorem \ref{thm:CSym3} for [P5] and Theorem \ref{thm:P6} for [P6], we obtain the following:
\begin{Thm}\label{thm:MAIN}
The state space $\SA$ of GPTs satisfying [P5] and [P6] with $\dim \SA =2,3$ is either simplex or unit disk.    
\end{Thm}
This result implies that GPTs with principles [P5] and [P6] such that the state space is embedded in 
$2$ or $3$ dimensional real vector space is restricted to either classical system or quantum system (qubit). 

Notice that this does not hold in higher dimensional GPTs as quantum systems with more than or equal to $3$ level systems satisfy both [P5] and [P6] while the state space is neither the simplex nor the ball \cite{ref:G}. 

\section{Conclusion and Discussion}\label{sec:CD}

In this paper, after confirming our basis of general GPTs (namely, Principles [P1]-[P4]), we have introduced and investigated two physical principles [P5] (Physical Equivalence of Pure States) and [P6] (Decomposability with Distinguishable Pure States). 
In particular, the former is established with the operationally motivated definition of physical equivalence of states, which turns out to be equivalent to the symmetric structure of the state space (See Sec.~\ref{subsec:PE}).  
In each principle, we gave a classification of state spaces (Theorems \ref{thm:CSym2}, \ref{thm:CSym3} for [P5] and \ref{thm:P6}) for [P6]) and by combining them, we showed that GPTs with $2$ or $3$ dimensional state spaces are either classical or quantum system (with the Bloch ball) (Theorem \ref{thm:MAIN}). 
Note that similar characterizations of the Bloch ball are derived in \cite{ref:Hardy,ref:DakicBrukner,ref:Masanes,ref:CDP}. 
However, the definition of the Bloch ball there is the state space of systems with at most two distinguishable states, while our derivation does not assume the maximum number of distinguishable states. 
In this paper, we focused on simple principles on states and measurements, and showed how each principle narrows down GPTs close to classical and quantum systems. 
In near future, we further consider other principles on especially correlations on composite systems, and discuss the characterization of QM by purely physical principles. 

\bigskip

{\bf Acknowledgment} 

\vspace{3mm}

We are grateful to Dr. Miyadera and Dr. Imafuku for their useful discussion.  
One of the authors (G.K.) would like to thank Prof. D'Ariano, Dr. Chiribella, and Dr. Perinotti for their fruitful discussion and comments at Nagoya Winter Workshop, 18-24 February (2010). 
This work was partially supported by Grant-in-Aid for Young Scientists (B) (No.20700017 and No.22740079), The Ministry of Education, Culture, Sports, Science and Technology (MEXT), Japan.

\appendix 

\section{}\label{app:Rep1}
In this appendix, we give a proof of Representation 1 which is formally described by the following Theorem:
\begin{Thm}\label{thm:BasicRep}
Assume [P1]-[P3] and let $\SA$ be the state space. 
Then, there exist a real vector space $V$ and a representation map $ \ \hat{} \ : \SA \to V$ 
such that (i) the map $\ \hat{} \ $ is injective (one-to-one) and $\widehat{\langle p; s_1,s_2 \rangle} = p \hat{s_1} + (1-p) \hat{s_2}$; (thus $\hat{\SA}$ is a convex subset of $V$
 and pure states correspond to extreme points of $\hat{\SA}$), and (ii) an $n$-valued measurement $M$ is represented by an $n$-tuple of effects $(e_i)_{i=1}^n$ on $\hat{\SA}$ such that 
$\mathrm{Pr}\{M = i | s \} = e_i(\hat{s})$ for any $s \in \SA$ (and thus $\sum_i e_i = u$ where $u$ is the unit effect). 
\end{Thm}
{\bf [Proof]} 
We call a real functional $f: \SA \to \R$ an ``affine'' functional iff $f(\langle p;s_1,s_2 \rangle) = p f(s_1) + (1-p) f(s_2)$ for any $s_1,s_2\in \SA$ and $p \in [0,1]$. 
Also an ``affine'' functional $e$ on $\SA$ is called an ``effect'' if the range is in $[0,1]$. 
The unit ``effect'' $u$ is the effect such that $u(s)= 1$ for any $s \in \SA$. 
We denote by $\A(\SA)$ and $\E(\SA)$ the set of all the ``affine'' functionals and ``effects'' on $\SA$. 
It is easy to see that $\A(\SA)$ forms a real vector space with pointwise addition and scalar multiplication, 
while $\E(\SA)$ is a convex subset of $\A(\SA)$. 


Let $M$ be an $n$-valued measurement. 
Then, the probability $\mathrm{Pr}\{M=i | s \} $ with a fixed $i$ is considered as a map from $\SA $ to $[0,1]$:
\begin{equation}\label{eq:ei}
s \in \SA \to e_i(s):= \mathrm{Pr}\{M=i | s \} \ (i=1,\ldots,n), 
\end{equation}
which are clearly ``effects'' from [P3-1]. 
Remind that measurements are characterized only through probabilities from [P2-2].  
Thus $M$ is characterized by $n$-tuple of ``effects'' $(e_i)_{i=1}^n$ where $e_i \in \E(\SA)$ is defined by \eqref{eq:ei}. 
From the normalization condition of the probability, we have $\sum_{i=1}^n e_i = u$. 
In the following, we denote by $M = (e_i)_{i=1}^n$ an $n$-valued measurement meaning that $e_i(s)$ is the probability to get $i$th outcome when performing $M$ under state $s$.  
Note that the part (ii) of Theorem \ref{thm:BasicRep} is an easy consequence of this representation of measurements. (See the last part of this proof.)

Next, let $V$ be the set of all the affine functionals on $\E(\SA)$ with bounded ranges. 
It is easy to see that $V$ is a real vector space with pointwise addition and scalar multiplication. 
Now we define a state-representation map $\hat{}: \SA \to V$ by 
\begin{equation}
\hat{s} (e) := e(s) \ (\forall e \in \E(\SA)). 
\end{equation} 
Note that $\hat{s}$ is an affine functional on $\E(\SA)$ by the definition, while the range of $\hat{s}$ is bounded since $0 \leq e(s) \leq 1$, thus $\hat{}$ maps any state to a vector in $V$. 
Let $\hat{s_1} = \hat{s_2} $ for $s_1,s_2 \in \SA$. Then $e(s_1) = \hat{s_1}(e) = \hat{s_2}(e) = e(s_2)$ for any ``effect'' $e$. 
Therefore, for any measurement $M = (e_i)_{i=1}^n$, $\mathrm{\Pr}\{M=m_i | s_1\} = e_i(s_1) = e_i(s_2) = \mathrm{Pr}\{M = m_i | s_2 \}$, which implies $s_1 = s_2$ from [P2-1]. 
Thus, $\hat{}$ is injective. 
From the ``affinity of effect'', $\widehat{\langle p;s_1,s_2\rangle} (e) = e (\langle p;s_1,s_2\rangle) = p e(s_1) + (1-p) e(s_2) = (p \hat{s_1} + (1-p) \hat{s_2}) (e) $ for any $e \in \E(\SA)$. 
Thus, the state $s = \langle p;s_1,s_2\rangle$ is mapped to be a convex combination of $\hat{s_1}$ and $\hat{s_2}$: 
\begin{equation}\label{eq:PM}
\widehat{\langle p;s_1,s_2\rangle} = p \hat{s_1} + (1-p) \hat{s_2}. 
\end{equation}
Notice that from [P3-1] the set $\hat{\SA}:= \{ \hat{s} \in V | s \in \SA \}$, which is the state-space of this representation, is a convex subset of $V$. 
It is straightforward from the definition that pure states are mapped to extreme points of $\hat{\SA}$, and vice versa. 
This completes the proof of part (i) of Theorem \ref{thm:BasicRep}. 

Finally, let $\check{}$ be a representation map on $\E(\SA)$ to the set of affine functionals on $\hat{\SA}$ defined by $\check{e}(\hat{s}) := e(s) \ (\forall s \in \SA)$. 
It is easy to see the injectivity and the affinity of $\ \check{}$. 
Since the range of $\check{e}$ for $e \in \E$ is in $[0,1]$, $\check{e}$ is an effect on $\hat{\SA}$. 
Therefore, any $n$-valued measurement $M$ is represented by an $n$-tuple of effects $(\check{e_i})_{i=1}^n$ on $\hat{\SA}$ such that $\sum_i \check{e_i}$ is the unit effect and 
$\mathrm{Pr}\{M = i | s \} = \check{e_i}(\hat{s})$ for any $s \in \SA$. 

\hfill $\blacksquare$

The representation symbols $\hat{}$ and $\check{} $ are all omitted throughout this paper except in the above proof. 



\section{``Physical Topology'' on State Space}\label{app:Rep2}

In this appendix, we review a physical topology and show Representation 2 in finite dimensional cases. 
For the readers' convenience and mathematical simplicity, we explain these using a Kolmogorov distance.   

Remind from Appendix \ref{app:Rep1} that each vector in $V$ is a bounded affine functional on $\E(\SA)$.
Then, one can introduce a natural norm on $V$ by  
\begin{equation}\label{eq:snorm}
||v|| := \sup_{e \in \E} |v(e)|. 
\end{equation}
(Notice that $||v|| < \infty$ as $v$ is a bounded functional on $\E$; it is clear that (i) $||v||\ge 0$; (ii) $||v||=0$ iff $v=0$; (iii) $||\alpha v|| = |\alpha| ||v||$; (iv) $||v + w || \le ||v|| + ||w||$ for all $v,w \in V$ and $\alpha \in \R$.) 
Notice that the norm of any state $s$ is $1$ as $|e(s)| \le 1$ for all $e \in \E$ and $u(s) = 1$. 
Thus, $\SA$ lies on the unit sphere with respect to this norm. 
The induced metric 
$$
D_{kol}(v,w):= ||v-w|| \ (v,w \in V)
$$
is called the Kolmogorov distance \cite{ref:FG}, since it is a straightforward generalization of the Kolmogorov distance in classical systems. 
In QM, the norm and distance are the trace norm and trace distance. 
The Kolmogorov distance on $\SA$ has an operational meaning through the relation  
\begin{equation}\label{eq:KolDis}
D_{kol}(s_1,s_2)=  2 P(s_1,s_2) - 1, 
\end{equation}
where $P(s_1,s_2)$ is the optimal success probability to distinguish states $s_1$ and $s_2$ which are prepared with probabilities $1/2$ and $1/2$ (see \cite{ref:KNI} and references therein). 
Therefore, the topology with respect to this norm has an operational meaning 
as states $s_1$ and $s_2$ are close w.r.t. this norm iff the optimal success probability to distinguish them are close to $1/2$.  
 
From \eqref{eq:KolDis}, state space $\SA$ is bounded in $V$ with this norm as probabilities are in $[0,1]$. 
In general, $\SA$ might not be closed. 
However, with a physical consideration, it turns out that without loss of generality one can assume $\SA$ is closed. 
Indeed, even if there exists an ``imaginary state'' $v \in V$ in the closure of $\SA$ which is not in $\SA$, 
there exists a state $s \in \SA$ arbitrary close to $v$ with this norm. 
Namely, for any small $\varepsilon > 0$, there exists a state $s$ such that $|v(e) - s(e)| < \varepsilon$ for any $e\in \E$. 
Thus, under the presence of an (arbitrary small but) finite error as always the cases in the real experiments, 
it is impossible to (even statistically) distinguish the ``imaginary state'' and state $s$ through all the possible measurements. 
Therefore, it is a matter of taste to include or not every ``imaginary states'' as states. 

To sum up, one can assume that $\SA$ is represented as a closed and bounded convex subset of a normed space $V$. 
Note that, if dim $V < \infty$, compactness and closed boundedness are equivalent from the Heine-Borel Theorem for general normed space \cite{ref:HB}. 
Moreover, in that case, this topology is the same as the physical topology \cite{ref:Araki} due to the uniqueness of the topology of the locally convex Hausdorff topological vector space.

We note the monotonicity holds \cite{ref:KNI} for any affine map $\Psi$ on $\SA$: 
\begin{equation}\label{eq:M}
||\Psi(s)|| \le ||s||. 
\end{equation}
and thus 
\begin{equation}\label{eq:MON}
||\Psi(s_1) - \Psi(s_2)|| \le ||s_1 - s_2|| \ (\forall s_1,s_2 \in \SA).  
\end{equation}
Applying this for an affine bijection on $\SA$, we get the following:  
\begin{Lem}\label{lem:InvNorm}
Any affine bijection $\Psi$ on $\SA$ preserves the Kolmogorov distance:  
$$
|| \Psi(s_1) - \Psi(s_2) || = ||s_1 - s_2||. 
$$
\end{Lem}
{\it Proof} Apply \eqref{eq:MON} for $\Psi$ and $\Psi^{-1}$. 
\hfill $\blacksquare$


\section{Extension of affine maps}\label{app:ExAff} 

Let $V_1,V_2$ be real vector spaces and $W$ be a convex subset of $V_1$. 
Then, any affine map $\Lambda:W \to V_2$ has an affine extension to $V_1$. 
With some additional conditions, $\Lambda$ is further extended to a linear map on $V_1$.    

In this appendix, we present the proof of these facts for finite dimensional cases \footnote{For infinite dimensional cases, by noting the existence of a basis of $V_1$ (guaranteed by Zorn's lemma), essentially the same proofs are applied. }: 
Let dim $V_1 < \infty$. 
Throughout this section, $\Lambda$ is an affine map from $W$ to $V_2$.

\begin{Lem}\label{lem:rest}
Let $w_i \in W $ and $\lambda_i \in \R \ (i=1,\ldots,m)$ with $\sum_i \lambda_i = 1$. 
If $\sum_{i=1}^m \lambda_i w_i \in W$, then 
\begin{equation}\label{eq:AEtoW}
\Lambda(\sum_{i=1}^m \lambda_i w_i) = \sum_{i=1}^m \lambda_i \Lambda(w_i). 
\end{equation}
\end{Lem}
{\it Proof}. 
Let $C_+:= \{i =\{1,\ldots,m\}| \lambda_i \ge 0\}$ and $C_-:= \{i =\{1,\ldots,m\}| \lambda_i < 0\}$. 
Since $\sum_{i=1}^m \lambda_i = 1$, it follows that $1 + \sum_{i \in C_-} |\lambda_i| = \sum_{i \in C_+} |\lambda_i|$, and thus 
$C_+ \neq \emptyset$, $\sum_{i \in C_+} |\lambda_i| =: \lambda \ge 1$, $|\lambda_i| \le \lambda $ for any $i=1,\ldots,m$. 
Letting $w:= \sum_{i=1}^m \lambda_i w_i \in W$, we have 
$$
\frac{1}{\lambda} w + \sum_{i\in C_-} \frac{|\lambda_i|}{\lambda} w_i = \sum_{i\in C_+} \frac{|\lambda_i|}{\lambda} w_i,  
$$
where both right and left sides are convex combinations of vectors from $W$. 
From the affinity of $\Lambda$, we have 
$$
\frac{1}{\lambda} \Lambda(w) + \sum_{i\in C_-} \frac{|\lambda_i|}{\lambda} \Lambda(w_i) = \sum_{i\in C_+} \frac{|\lambda_i|}{\lambda} \Lambda(w_i),
$$
which implies \eqref{eq:AEtoW}. 
\hfill $\blacksquare$

\begin{Prop}\label{prop:AE}
$\Lambda$ has an affine extension $\tilde{\Lambda}$ to $V_1$ with the form 
\begin{equation}\label{eq:exLam}
\tilde{\Lambda}(v) = A v + b
\end{equation}
where $A$ is a linear map from $V_1$ to $V_2$ and $b \in V_2$. 
If $\mathrm{Aff} (W) = V_1$, then the extension is unique. 
 \end{Prop}
{\it Proof.} Let dim Aff$(W) = n$ and dim $V_1 = n + m \ ( m \ge 0)$. 
Let $w_i \in W \ (i=0,\ldots,n)$ be affinely independent set so that $w_i - w_0 \ (i=1,\ldots,n)$ are linearly independent. 
Let $\{v_i\}_{i=1}^{n+m}$ be a basis of $V_1$ such that $v_i = w_i - w_0 $ for $i=1,\ldots,n$. 
Then, define a map $\tilde{\Lambda}:V_1 \to V_2$ in \eqref{eq:exLam} with a linear map $A:V_1 \to V_2$ and $b \in V_2$ given by    
\begin{eqnarray*}
A v_i &:=& \left\{
\begin{array}{cc}
\Lambda(w_i) - \Lambda(w_0) & i=1,\ldots,n\\
v'_i & i= n+1,\ldots,n+m  
\end{array} 
\right. \\
b &:=& \Lambda(w_0) - A w_0 
\end{eqnarray*}
where $v'_i$ are arbitrary vectors in $V_2$. 
Then, one can show that $\tilde{\Lambda}$ is an affine extension of $\Lambda$. 
To see this, note first that any $w \in W$ is uniquely written by $w = \sum_{i=0}^n \lambda_i w_i$ with $\lambda_i \in \R$ such that $\sum_{i=0}^n \lambda_i = 1$ since 
$\{w_i\}_{i=0}^n$ are affinely independent and $n$ is the dimension of $W$. 
Then, for any $w = \sum_{i=0}^n \lambda_i w_i \in W$, it follows that $\tilde{\Lambda}(w) = (\sum_{i=1}^n \lambda_i Aw_i) + \lambda_0 A w_0 + (\Lambda(w_0) - A w_0) 
= \sum_{i=1}^n \lambda_i (A w_0 + \Lambda (w_i) - \Lambda(w_0)) + \lambda_0 A w_0 + (\Lambda(w_0) - A w_0)
= \sum_{i=0}^n \lambda_i \Lambda (w_i) - \lambda_0 \Lambda(w_0) + (1-\lambda_0) (A w_0 -\Lambda(w_0)) + \lambda_0 A w_0 + (\Lambda(w_0) - A w_0)
 = \Lambda(w)$, where we have used Lemma \ref{lem:rest} in the final equation. 
Affinity of $\tilde{\Lambda}$ follows from the form of \eqref{eq:exLam}.

Let Aff$(W) = V_1$. 
Assume that $\tilde{\Lambda}$ and $\tilde{\Lambda}'$ are affine extensions of $\Lambda$. 
Noting that any vector $v \in V_1 = \mathrm{Aff}(W)$ is uniquely written by $v = \sum_{i=0}^n \lambda_i w_i$ with $\sum_{i=0}^n \lambda_i = 1$, 
we have $\tilde{\Lambda}(v) = \sum_{i=0}^n \lambda_i \tilde{\Lambda}( w_i ) = \sum_{i=0}^n \lambda_i \Lambda( w_i ) = \sum_{i=0}^n \lambda_i \tilde{\Lambda}'( w_i ) = \tilde{\Lambda}'(v)$ by the same argument as Lemma \ref{lem:rest} (note that each $w_i$ is in the domain $W$ of $\Lambda$). 
Thus the uniqueness of the extension follows. 
\hfill $\blacksquare$

The following result gives a linear extension of $\Lambda$:  
\begin{Prop}\label{prop:LExt}
(i) If $0 \not \in \mathrm{Aff}(W)$, $\Lambda$ has always a linear extension to $V_1$. 
In particular, if dim $V_1$ = $\mathrm{dim \ Aff}(W)$ + 1, the extension is unique. 
(ii) If $0 \in \mathrm{Aff}(W)$, then $\Lambda$ does not necessarily have linear extensions. 
In particular, in the case $0 \in W$, $\Lambda$ has a linear extension to $V_1$ iff $\Lambda(0) = 0$. 
\end{Prop}
{\it Proof}. 
Here we use the same notation as in the proof of Proposition \ref{prop:AE}. 

(i) Let $0 \not \in \mathrm{Aff}(W)$. 
Obviously, Aff$(W) \neq V_1$ and $m \ge 1$. 
Using the basis $\{v_i\}_{i=1}^{n+m}$, let $- w_0 = \sum_{i=1}^{n+m} x_i v_i$ with some $x_i \in \R$. 
Then, there exists at least one non-zero coefficient $x_i \ (i=n+1,\ldots,m)$; Otherwise we have $-w_0 = \sum_{i=1}^n x_i(w_i-w_0)$ and thus $0 = \sum_{i=1}^n x_i w_i + (1- \sum_{i=1}^n x_i) w_0 \in \mathrm{Aff}(W)$, which contradicts $0 \not\in \mathrm{Aff}(W)$. 
Without loss of generality, let $x_{n+1} \neq 0$ and choose $v^\prime_{n+1} = - \frac{1}{x_{n+1}}(\Lambda(w_0) +\sum_{i=1}^n x_i (\Lambda(w_i) - \Lambda(w_0)) ) $ and 
$v'_i = 0 \ (i=n+2,\ldots,n+m)$. 
Then, a direct computation shows that $A w_0 = \Lambda(w_0)$, i.e., $b = 0$. 

Next, we show the uniqueness of the linear extension for the case dim $V$ = dim Aff$(W) + 1$, i.e., $m = 1$. 
In this case, one can choose $v_{n+1} := w_0$ in a basis $\{v_i\}_{i=1}^{n+1}$.  
Indeed, if $\sum_{i=1}^n \alpha_i (w_i - w_0) + \alpha_{n+1} w_0 = 0$ and at least one $\alpha_i$ is non-zero, then $\alpha_{n+1} \neq 0$ by linear independence of $w_i - w_0 = v_i$ ($1 \leq i \leq n$), but this implies $0 \in $Aff$(W)$, which is a contradiction.  
Therefore, $\{v_i\}_{i=1}^n$ with $v_i = w_i-w_0$ and $v_{n+1}= w_0$ is an independent system of $V$ and thus gives its basis. 
Let $A$ and $A'$ give linear extensions of $\Lambda$. 
Then, using the above basis, any vector $v \in V$ is written as $v = \sum_i x_i (w_i-w_0) + x_{n+1} w_0$. 
Since $A w = \Lambda(w) = A'w$ for any $w \in W$, we have $A v = A' v$ for any $v \in V$. 
This proves the uniqueness of the extension. 

(ii) Let $0 \in W$. 
Then $b = \Lambda(0)$. 
Therefore, in this case, an affine map can be linearly extended iff $\Lambda(0) = 0$. 
\hfill $\blacksquare$

It is well-known that a quantum operation $\Lambda$ on the set $\SA_q$ of density operators is uniquely extended to a linear map on a vector space ${\cal L}(\HA_d)$ of all the linear operators on a Hilbert space $\HA_d$. 
While there are several ways to show this extension, it can be derived from Proposition \ref{prop:LExt} as an easy corollary. 
Indeed, it is easy to check that $\SA_q$ is a convex subset of a real vector space $V_1 := \LA(\HA)_h$ of the set of Hermitian operators, satisfying 
$0 \not \in \mathrm{Aff}(\SA_q)$ and dim $\LA(\HA)_h$ = dim Aff$(\SA_q) + 1$. 
Therefore, one can uniquely extend $\Lambda$ to a linear map $\tilde{\Lambda}$ on $\LA(\HA)_h$. 
Next, noting that any linear operator $A \in \LA(\HA)$ is uniquely decomposed into the sum of real and imaginary parts, i.e., $A = A_R + iA_I$ where $A_R := \frac{1}{2}(A + A^\dagger)$ and 
$A_I:= \frac{1}{2i}(A - A^\dagger)$, one can further extend $\tilde{\Lambda}$ to a linear map $\tilde{\tilde{\Lambda}}$ on $\LA(\HA)$ by $\tilde{\tilde{\Lambda}}(A):= \tilde{\Lambda}(A_R) + i \tilde{\Lambda}(A_I)$. 

We notice that, in general GPTs, one can also treat operations (including dynamical map and measurement processes) described by its linear extension in a similar manner.

In the following, we assume $\mathrm{Aff}(W) = V_1 = V_2$ and let $\Lambda : W \to W $ be an affine map on $W$. 
We show that any surjective affine map on $W$ is affinely extended to a bijective map on $V := V_1$. 
\begin{Lem}
If $\Lambda$ is surjective on $W$, so is the affine extension $\tilde{\Lambda}:V \to V$ on $V$.  
\end{Lem}
{\it Proof} \ As $\mathrm{Aff}(W) = V$, an arbitrary $y \in V$ is written as $y = \sum_{i=0}^n \lambda_i w_i$ with some $\lambda_i \in \R$ and $\sum_i \lambda_i = 1$. 
From the surjectivity of $\Lambda$ on $W$, there exists $w_i^\prime$ such that $w_i = \Lambda(w_i^\prime) \ (\forall i = 0,\ldots,n)$. 
Therefore, we have $y = \sum_i \lambda_i \Lambda(w_i^\prime) = \sum_i \lambda_i (A w_i^\prime + b) = A (\sum_i \lambda_i w_i^\prime) + b = \tilde{\Lambda}(\sum_i \lambda_i w_i^\prime)$. 
Thus, $\tilde{\Lambda}$ is surjective on $V$.  
\hfill $\blacksquare$ 
\begin{Prop}\label{Prop:StoB}
If $\Lambda$ is surjective on $W$, then $\tilde{\Lambda}:V \to V$ and $\Lambda$ are bijective.  
\end{Prop}
{\it Proof} \ From the previous Lemma, $\tilde{\Lambda}$ is surjective on $V$ and so is the linear map $A: V \to V$ which composes $\Lambda$. 
Remind that a linear map on a finite dimensional vector space is surjective iff it is injective.  
Therefore, $A$ is bijective and so is $\tilde{\Lambda}$ on $V$. 
Injectivity of $\Lambda$ on $W$ follows since it is a restriction of the injective map $\tilde{\Lambda}$ to $W$.  
\hfill $\blacksquare$


\section{}\label{sec:A}
In this appendix, we give proofs of Lemma \ref{lem:bPhi} and Theorem \ref{thm:AF}. 

\noindent {\bf [Proof of Lemma \ref{lem:bPhi}]} \ ($1 \Rightarrow 2$) If $e \leq f$, then $g := f/2$ and $k := f - e $ are clearly effects.
As $g = (f + 0)/2 = (e + k)/2 \in \mathcal{E}$, we have $\Phi(g) = (\Phi(f) + 
\Phi(0))/2 = (\Phi(e) + \Phi(k))/2$. 
Since $\Phi(0) = 0$, we have $\Phi(f) = \Phi(e) + \Phi(k)$, and thus 
$\Phi(e) \leq \Phi(f)$. ($2 \Rightarrow 3$) By the relation $\Phi^{-1}(u) \leq u$ and the assumption, we have $u = \Phi(\Phi^{-1}(u)) \leq \Phi(u)$, therefore $\Phi(u) = u$.
($3 \Rightarrow 1$) For any $g \in \E$, $u-g \in \E$ and thus $(\Phi(g)+\Phi(u-g))/2 = (\Phi(u)+\Phi(0))/2 \in \E$. 
Since $\Phi(u)=u$, we get $\Phi(g)+\Phi(u-g) - \Phi(0) = u \ge \Phi(u-g)$, therefore $\Phi(g) \ge \Phi(0)$. 
Put $g = \Phi^{-1}(0)$, we have $0 \ge \Phi(0)$, and thus $0 = \Phi(0)$. 
\hfill $\blacksquare$

Let $\SA \subset \R^d$ with Aff $(\SA) = \R^d$, then from Proposition \ref{prop:AE}, any effect $e \in \E$ is uniquely represented by $a\in \R^{d},b \in \R$ such that 
$$
e(s) = \langle a, s \rangle + b,
$$
where $\langle \cdot, \cdot \rangle$ represents the Euclidean inner product in $\R^n$. 
(Remind that any real linear functional $A:\R^d \to \R$ is uniquely represented by a vector $a \in \R^n$ such that $A x = \langle a, x \rangle$ for any $x \in \R^d$.) 
Let a map $\tilde{}:\E \to \R^{d+1}$ be defined by $\tilde{e} = (a,b)$, then it is easy to see that $\tilde{}$ is an affine injection and 
$\widetilde{\E}$ is a convex subset in $\R^{d+1}$. 
In particular, $\tilde{u} = (0,1)$ and $\tilde{0} = (0,0)$.  
 
\noindent {\bf [Proof of Theorem \ref{thm:AF}] } Let $\tilde{\Lambda}$ be an affine functional on $\widetilde{\E}$ defined by $\tilde{\Lambda} (\tilde{e}):= \Lambda(e)$. 
Note that $(0,0) = \tilde{0} \in \tilde{\E}$ and $\tilde{\Lambda}(0,0) = \Lambda(0) = 0$. 
Thus, from Proposition \ref{prop:LExt}, $\tilde{\Lambda}$ can be represented by a vector $(x,y) \in \R^d \times \R =\R^{d+1}$ such that 
$$
\tilde{\Lambda}(a,b) = \langle (x,y) , (a,b) \rangle = \langle x, a\rangle + yb.
$$
Since $1 = \Lambda(u) = \tilde{\Lambda}(0,1)$, we have $y = 1$:
$$
\tilde{\Lambda}(a,b)  = \langle x,a\rangle  + b.
$$
In the following, we prove that $s: = x \in \SA$, and thus for any effect $e \in \E$ with $\tilde{e} =: (a,b)$, we have 
$\Lambda(e) = \tilde{\Lambda}(a,b) = \langle s, a \rangle + b = e(s)$.  
 
To prove this, assume contrary that $x \not\in \SA$, 
then there exists a hyperplane which strictly separates $\{x \}$ and $\SA$ (see Theorem 4.12 in \cite{ref:Lay});  
Equivalently, there exists a (continuous) affine functional $f:\R^d \to \R$ such that $f(x)<0$ and $f(t) > 0$ for any $t \in \SA$. 
Then, $f':= \frac{1}{\max_{t\in \SA}f(t)}f $ is an effect such that $0> f'(x)$. 
Note that $f'$ can be represented as $f'(z) = \langle a_0,z \rangle + b_0 \ (z \in \R^d)$, and thus $\tilde{f'} = (a_0,b_0)$ from the uniqueness of the representation of effects. 
Thus, we have 
$$
0 > f'(x) = \langle a_0,x \rangle + b_0 = \tilde{\Lambda}(a_0,b_0) = \Lambda{(f')}. 
$$ 
This contradicts that $\Lambda (e) \ge 0$ for any $e \in \E$.  
 
Finally, the uniqueness follows from the separation hypothesis for states [P2-1]. 
\hfill $\blacksquare$

\end{document}